\documentclass[prd,nofootinbib,preprint,showpacs,showkeys,superscriptaddress]{revtex4}
\usepackage[english]{babel}
\usepackage{amsmath,amssymb}
\usepackage[dvips]{graphicx}
\usepackage{ifthen,array}
\usepackage[sort&compress]{natbib}
\usepackage{amsfonts}
\usepackage{bbm}
 
\newcommand{\be}{\begin{equation}}
\newcommand{\ee}{\end{equation}}
\newcommand{\bea}{\begin{eqnarray}}
\newcommand{\eea}{\end{eqnarray}}

\newcommand{\lsim}{
\mathrel{\hbox{\rlap{\hbox{\lower4pt\hbox{$\sim$}}}\hbox{$<$}}}}
\newcommand{\gsim}{
\mathrel{\hbox{\rlap{\hbox{\lower4pt\hbox{$\sim$}}}\hbox{$>$}}}}

\let\vev\VEV
\def\e6{$E(6)$}
\def\10{$SO(10)$}
\def\21{$SU(2) \otimes U(1) $}

\def\422{$SU(4) \otimes SU(2) \otimes SU(2)$}
\def\321{$SU(3) \otimes SU(2) \otimes U(1)$}
\def\lsim{\raise0.3ex\hbox{$\;<$\kern-0.75em\raise-1.1ex\hbox{$\sim\;$}}}
\def\gsim{\raise0.3ex\hbox{$\;>$\kern-0.75em\raise-1.1ex\hbox{$\sim\;$}}}

\newcommand{\ed}{\end{document}}
\DeclareMathAlphabet{\mathsc}{OT1}{cmr}{m}{sc}

\newcommand{\CL}   {C.L.}
\newcommand{\dof}  {d.o.f.}

\newcommand{\eVq}  {\rm{eV}^2}
\newcommand{\Sol}  {\mathsc{sol}}
\newcommand{\Atm}  {\mathsc{atm}}

\newcommand{\Dms}  {\Delta m^2_\Sol}
\newcommand{\Dma}  {\Delta m^2_\Atm}

\def \nbb {$\beta\beta_{0\nu}$ }

\renewcommand{\baselinestretch}{1.45}

\newcommand{\AddrAHEP}{%
 AHEP Group, Instituto de F\'{\i}sica Corpuscular,
  C.S.I.C. -- Universitat de Val{\`e}ncia \\
  Edificio de Institutos de Paterna, Apartado 22085,
  E--46071 Val{\`e}ncia, Spain\\}

\begin{document}

\title{A Theory Perspective on Neutrino Oscillations}

\author{J. W. F. Valle}
\vskip 2cm
\affiliation{\AddrAHEP} 

\vskip 2cm

\begin{abstract}
  
  I summarize the status of neutrino oscillations that follow
  from current data, including the status of the small parameters 
  $\alpha \equiv \Dms/\Dma$ and $\sin^2\theta_{13}$ characterizing
  the strength of CP violation in neutrino oscillations. I briefly
  discuss the impact of oscillation data on the prospects for probing the 
  absolute scale of neutrino mass in neutrinoless double beta decay. 
  I also comment on the theoretical origin of neutrino mass, mentioning 
  recent attemps to explain current data from first principles.

\end{abstract}

\maketitle


\section{Introduction}

Nowadays neutrino physics lies at the center of attention of the particle, nuclear and astrophysics communities. 
Here I summarize the determination of neutrino mass and mixing
parameters from current neutrino experiments, given in detail in Ref.~\cite{Maltoni:2004ei}. (Waiting for the veredictum of MiniBoone, we neglect the LSND data). 
The key concept in these neutrino oscillation studies is that of neutrino mixing, 
a characteristic feature of gauge theories of massive neutrinos. 
The most complete study of the structure of the neutrino mixing matrix was 
given in~\cite{schechter:1980gr}.
For the analysis of current neutrino oscillation data  one assumes its simplest 
unitary, 3-by-3, and CP conserving form, as there is currently no sensitivity to 
CP violation.
The interpretation of the data requires good calculations of solar and
atmospheric neutrino fluxes~\cite{Bahcall:2004fg,Honda:2004yz},
neutrino cross sections and response functions, as well
the inclusion of matter effects~\cite{mikheev:1985gs,wolfenstein:1978ue} 
in the Sun and the Earth.
The reader is referred to Ref.~\cite{Maltoni:2004ei} for technical details 
and further references. For the discovery prospects of future neutrino 
oscillation studies see the Neutrino Oscillation Industry Web-Page~\cite{industry}.
Testing for the effect of leptonic CP phases and the absolute scale
of neutrino mass constitutes the main upcoming challenge.
Dirac CP phases will be probed in future oscillations studies, while 
Majorana phases will be tested in future searches for \nbb (neutrinoless double 
beta decay). 
I also briefly discuss the robustness of the oscillation hypothesis itself
{\sl vis a vis} the presence of non-standard physics. 

Last, but no least, where does the neutrino mass come from? I briefly comment on two 
alternative views on the theoretical origin of neutrino mass, mentioning some 
of their possible signatures.
                                                                     
\section{Solar and reactor data}
\label{sec:solar-+-kamland}

The solar neutrino data includes the rates of the chlorine experiment
($2.56 \pm 0.16 \pm 0.16$~SNU), the gallium results of SAGE 
($66.9~^{+3.9}_{-3.8}~^{+3.6}_{-3.2}$~SNU) and GALLEX/GNO ($69.3
\pm 4.1 \pm 3.6$~SNU), as well as the 1496--day Super-K data (44 bins:
8 energy bins, 6 of which are further divided into 7 zenith angle
bins). The SNO sample includes the data from the salt phase in the form
of the neutral current (NC), charged current (CC) and elastic
scattering (ES) fluxes, the 2002 spectral day/night data (17 energy
bins for each day and night period) and the 391--day data.  The
analysis includes both statistical errors, and systematic
uncertainties such as those of the eight solar neutrino fluxes.

Reactor anti-neutrinos from KamLAND are detected at the Kamiokande site 
by the process $\bar\nu_e + p \to e^+ + n$, where the delayed coincidence 
of the prompt energy from the positron and a characteristic gamma from
the neutron capture allows an efficient reduction of backgrounds.
Most of the incident $\bar{\nu}_e$'s come from nuclear plants at
distances of $80-350$ km from the detector, far enough to probe large
mixing angle (LMA) oscillations.
To avoid large uncertainties associated with geo-neutrinos an energy 
cut at 2.6~MeV prompt energy is applied for the oscillation analysis.

The first KamLAND data corresponding to a 162 ton-year exposure gave 54
anti-neutrino events in the final sample, after cuts, whereas $86.8
\pm 5.6$ events are predicted for no oscillations with $0.95\pm 0.99$
background events. This is consistent with the no--disappearance hypothesis 
at less than 0.05\% probability, giving the first evidence for the
disappearance of reactor neutrinos before reaching the detector, and
thus the first terrestrial confirmation of oscillations with $\Dms$.
                      
Additional KamLAND data with a larger fiducial volume of the
detector were presented at Neutrino 2004, corresponding to an
766.3~ton-year exposure.  In total 258 events have been observed,
versus $356.2\pm 23.7$ reactor neutrino events expected in the case of
no disappearance and $7.5\pm 1.3$ background events. This leads to a
confidence level of 99.995\% for $\bar\nu_e$ disappearance.  Moreover
they obtain evidence for spectral distortion consistent with oscillations.

A very convenient way to bin the latest KamLAND data is in terms of
$1/E_\mathrm{pr}$, rather than $E_\mathrm{pr}$. Various systematic errors 
associated to the neutrino fluxes, backgrounds, reactor fuel composition and 
individual reactor powers, small matter effects, and improved $\bar{\nu}_e$ flux
parameterization are included~\cite{Maltoni:2004ei}.  Assuming CPT invariance one 
can directly compare the information obtained from solar neutrino experiments with 
the KamLAND reactor results.

The KamLAND data single out the LMA solution from the previous ``zoo'' 
of alternatives~\cite{Maltoni:2003da}.  The stronger evidence for
spectral distortion in these data also leads to improved $\Dms$
determination, substantially reducing the allowed region of
oscillation parameters.  More than just cornering the oscillation
parameters, however, KamLAND has eliminated all previously viable 
non-oscillation solutions. From this point of view KamLAND has 
played a key role in the resolution of the solar neutrino problem.

\section{Atmospheric and accelerator data}
\label{sec:atmospheric-+-k2k}
                                                                               
The first evidence for neutrino conversions was the zenith angle
dependence of the $\mu$-like atmospheric neutrino data from the
Super-K experiment in 1998, an effect also seen in other atmospheric
neutrino experiments.  At the time there were equally good 
non-oscillation solutions, involving non-standard interactions~\cite{Gonzalez-Garcia:1998hj}. 
Thanks to the accumulation of upgoing muon data, and the observation of the dip in
the $L/E$ distribution of the atmospheric $\nu_\mu$ survival
probability, the signature for atmospheric neutrino oscillations has
now become convincing.  The data include Super-K charged-current atmospheric
neutrino events, with the $e$-like and $\mu$-like data samples of sub-
and multi-GeV contained events grouped into 10 zenith-angle bins, with
5 angular bins of stopping muons and 10 through-going bins of up-going
muons.  We do not use $\nu_\tau$ appearance, multi-ring $\mu$ and
neutral-current events, since an efficient Monte-Carlo simulation of
these data would require further details of the Super-K experiment, in
particular of the way the neutral-current signal is extracted from the
data. As far as atmospheric neutrino fluxes are concerned, we employ 
the latest three--dimensional calculations given in ~\cite{Honda:2004yz}.

The disappearance of $\nu_\mu$'s over a long-baseline probing the same 
$\Delta m^2$ region relevant for atmospheric neutrinos is now available 
from the KEK to Kamioka (K2K) neutrino oscillation experiment.
Neutrinos produced by a 12~GeV proton beam from the KEK proton
synchrotron consist of 98\% muon neutrinos with a mean energy of
1.3~GeV. The beam is controlled by a near detector 300~m away from the
proton target.  Comparing these near detector data with the $\nu_\mu$
content of the beam observed by the Super-K detector at a distance of
250~km gives information on neutrino oscillations.

The first phase (K2K-I data sample, corresponding to 
$4.8\times 10^{19}$ protons on target) gave 56 events in Super-K, 
whereas $80.1^{+6.2}_{-5.4}$ were expected for no oscillations. 
The second phase (K2K-II data, corresponding to $4.1\times 10^{19}$
protons on target) gave 108 events in Super-K, to be compared with
$150.9^{+11.6}_{-10.0}$ expected for no oscillations. Out of the 108
events 56 are so-called single-ring muon events.  This data sample
contains mainly muon events from the quasi-elastic scattering $\nu_\mu
+ p \to \mu + n$, and the reconstructed energy is closely related to
the true neutrino energy.  The K2K collaboration also finds that the
observed spectrum is consistent with the one expected for no
oscillation only at a probability of 0.11\%, whereas the best fit
oscillation hypothesis spectrum has a probability of 52\%.

One finds that the neutrino mass-squared difference inferred from the
$\nu_\mu$ disappearance in K2K agrees with atmospheric neutrino
results, providing the first terrestrial confirmation of oscillations with 
$\Dma$ with accelerator neutrinos. Due to low statistics the current K2K 
data sample gives a rather weak constraint on the mixing angle. 
However, although the determination of $\sin^2\theta_\Atm$ is
completely dominated by atmospheric data, K2K data already start
constraining the allowed $\Dma$ values~\cite{Maltoni:2004ei}.
For example, there is a constraint on $\Dma$ from below, which is
important for future long-baseline experiments, since their 
sensitivities are drastically affected if $\Dma$ lies in the lower 
part of the 3$\sigma$ range indicated by current atmospheric data.
                                                                               
\section{Three-neutrino oscillation parameters}
 
Lepton mixing is a characteristic feature of gauge theories of 
massive neutrinos. The first systematic study of the effective form
of the lepton mixing matrix was given in~\cite{schechter:1980gr}.  
For 3-neutrinos the simplest form of this matrix can be taken as 
\begin{equation}
  \label{eq:2227}
K =  \omega_{23} \omega_{13} \omega_{12}
\end{equation}
where each factor in the product of the $\omega$'s is effectively $2\times 2$,
characterized by an angle and a CP phase. Two of the three angles 
are involved in solar and atmospheric oscillations, so we set 
$\theta_{12} \equiv \theta_\Sol$ and $\theta_{23} \equiv \theta_\Atm$.  
The last angle in the three--neutrino leptonic mixing matrix is $\theta_{13}$, 
$$\omega_{13} = \left(\begin{array}{ccccc}
c_{13} & 0 & e^{i \phi_{13}} s_{13} \\
0 & 1 & 0 \\
-e^{-i \phi_{13}} s_{13} & 0 & c_{13}
\end{array}\right)\,.
$$
for which only an upper bound currently exists.  All three phases
are physical~\cite{schechter:1981gk}.  Two of the phases are fundamental,
and arise at the two-generation level, being associated to the Majorana nature 
of neutrinos. They show up only in lepton-number violating processes, like 
neutrinoless double beta decay, not in conventional neutrino oscillations~\cite{schechter:1981gk,doi:1981yb}. The other phase corresponds to the phase
present in the quark sector (Dirac-phase) and exists only with three
(or more) neutrinos. This phase affects neutrino oscillations. 

Such unitary form for the lepton mixing matrix holds in models where 
neutrino masses arise in the absence of right-handed neutrinos. To a 
good approximation, it also holds if neutrino masses are induced by 
a high-energy-scale seesaw mechanism (see below). 

In our analysis we follow the simplest unitary form in Eq. \ref{eq:2227}.
Since current neutrino oscillation experiments are insensitive to CP violation, 
we also neglect all phases in the analysis. In this approximation oscillations depend 
on the three mixing parameters $\sin^2\theta_{12}, \sin^2\theta_{23}, \sin^2\theta_{13}$ 
and on the two mass-squared splittings $\Dms \equiv \Delta m^2_{21} \equiv m^2_2
- m^2_1$ and $\Dma \equiv \Delta m^2_{31} \equiv m^2_3 - m^2_1$
characterizing solar and atmospheric neutrinos.  The hierarchy $\Dms
\ll \Dma$ implies that one can set $\Dms = 0$, to a good approximation, 
in the analysis of atmospheric and K2K data. Similarly, one can set $\Dma$ 
to infinity in the analysis of solar and KamLAND data.
Apart from the data already mentioned, the analysis also includes the 
constraints from "negative" reactor experiments, CHOOZ and Palo Verde.
                                                                               
\begin{figure}[t] \centering
    \includegraphics[width=.95\linewidth]{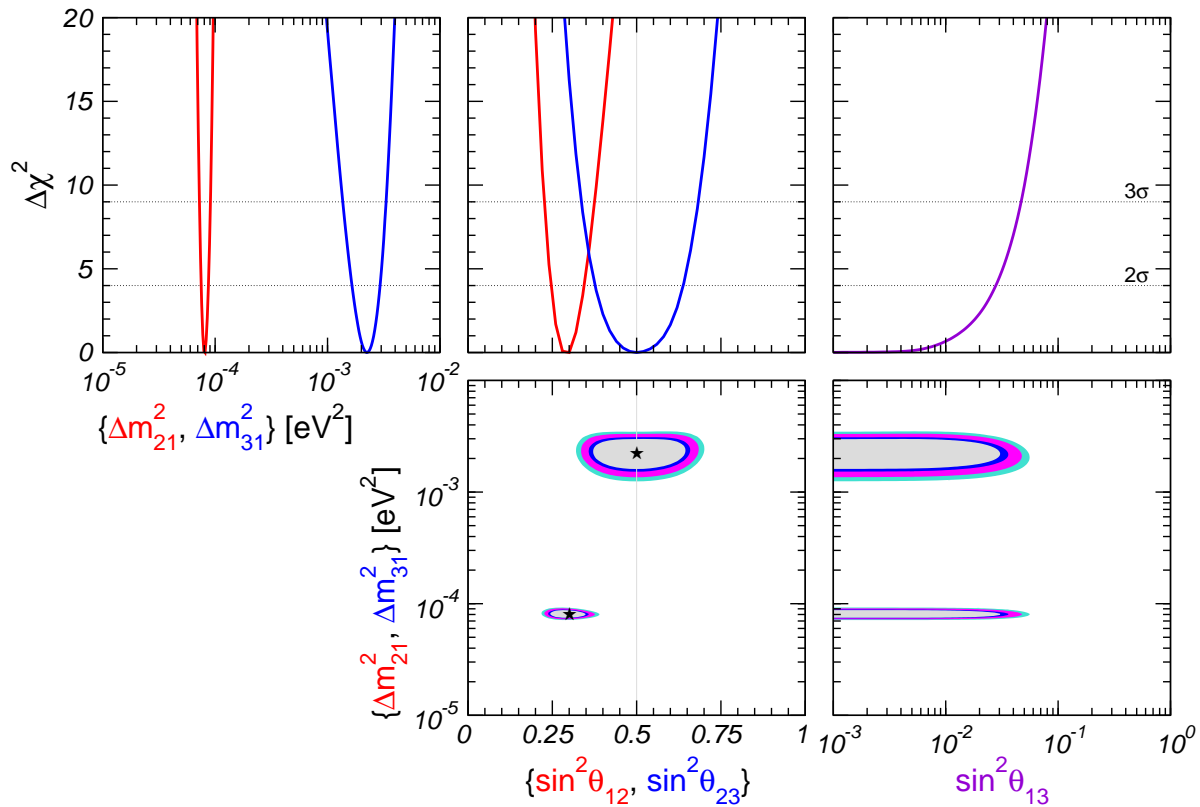}
    \caption{\label{fig:global} %
      Three--neutrino regions allowed by the world's neutrino oscillation data
      at 90\%, 95\%, 99\%, and 3$\sigma$ \CL\ for 2 \dof\ In top
      panels $\Delta \chi^2$ is minimized wrt undisplayed parameters.}
\end{figure}

The three--neutrino oscillation parameters that follow from the global 
oscillation analysis in Ref.~\cite{Maltoni:2004ei} are summarized in 
Fig.~\ref{fig:global} and in Table~\ref{tab:summary}.
In the upper panels of the figure the $\Delta \chi^2$ is shown as a function 
of the parameters $\sin^2\theta_{12}, \sin^2\theta_{23}, \sin^2\theta_{13}, \Delta
m^2_{21}, \Delta m^2_{31}$, minimized with respect to the undisplayed
parameters. The lower panels show two-dimensional projections of the
allowed regions in the five-dimensional parameter space. The best fit
values and the allowed 3$\sigma$ ranges of the oscillation parameters
from the global data are summarized in Table~\ref{tab:summary}.  This
table gives the current status of neutrino oscillation parameters.
\begin{table}[t] \centering    \catcode`?=\active \def?{\hphantom{0}}
      \begin{tabular}{|l|c|c|}        \hline        parameter & best
      fit & 3$\sigma$ range         \\  \hline\hline        $\Delta
      m^2_{21}\: [10^{-5}~\eVq]$        & 7.9?? & 7.1--8.9 \\
      $\Delta m^2_{31}\: [10^{-3}~\eVq]$        & 2.2?? &  1.4--3.3 \\
      $\sin^2\theta_{12}$        & 0.30? & 0.24--0.40 \\
      $\sin^2\theta_{23}$        & 0.50? & 0.34--0.68 \\
      $\sin^2\theta_{13}$        & 0.000 & $\leq$ 0.043 \\
      \hline
\end{tabular}    \vspace{2mm}
\caption{\label{tab:summary} Neutrino oscillation parameters~\cite{Maltoni:2004ei}.}
\end{table}

Note that in a three--neutrino scheme CP violation disappears when two 
neutrinos become degenerate~\cite{schechter:1980gr} or when one angle 
vanihes~\cite{schechter:1980bn}.  As a result its effects involve both the 
small mass hierarchy parameter $\alpha \equiv \Dms/\Dma$ as well as 
the small mixing angle $\theta_{13}$.
\begin{figure}[t]
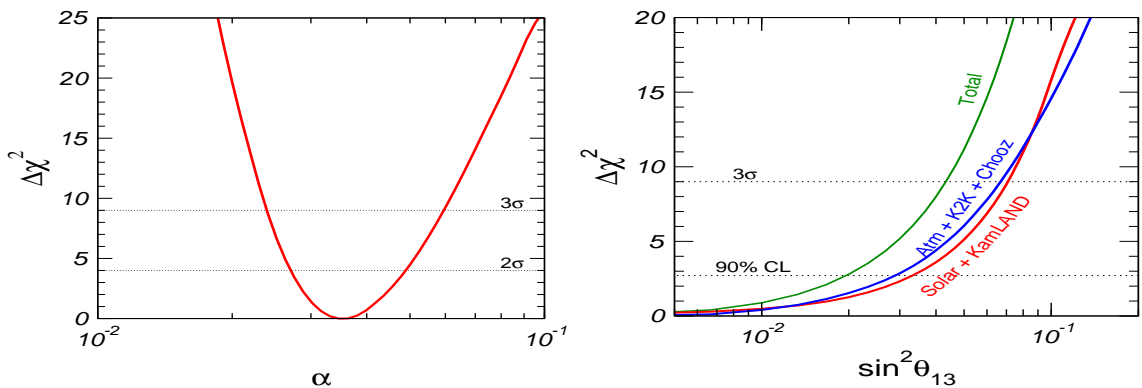
 \centering
  \includegraphics[height=5cm,width=.45\linewidth]{F-fcn.alpha-2.eps}
\includegraphics[height=5cm,width=.45\linewidth]{F-th13-chisq05.eps}
        \caption{\label{fig:alpha}%
      Determination of $\alpha \equiv \Dms / \Dma$ and bound on
      $\sin^2\theta_{13}$ from current data, from \cite{Maltoni:2004ei}.}
\end{figure}
The left panel in Fig.~\ref{fig:alpha} gives the parameter $\alpha$,
namely the ratio of solar over atmospheric splittings, as determined
from the global $\chi^2$ analysis.
The right panel in Fig.~\ref{fig:alpha} gives $\Delta\chi^2$ as a
function of $\sin^2\theta_{13}$ for different data samples.  One finds
that the KamLAND-2004 data have a surprisingly strong impact on this
bound. Before KamLAND-2004 the overall bound on $\sin^2\theta_{13}$ 
was dominated by the CHOOZ reactor experiment, together with the 
determination of $\Delta m^2_{31}$ from atmospheric data.

\begin{figure}[t] \centering
    \includegraphics[height=7cm,width=.7\linewidth]{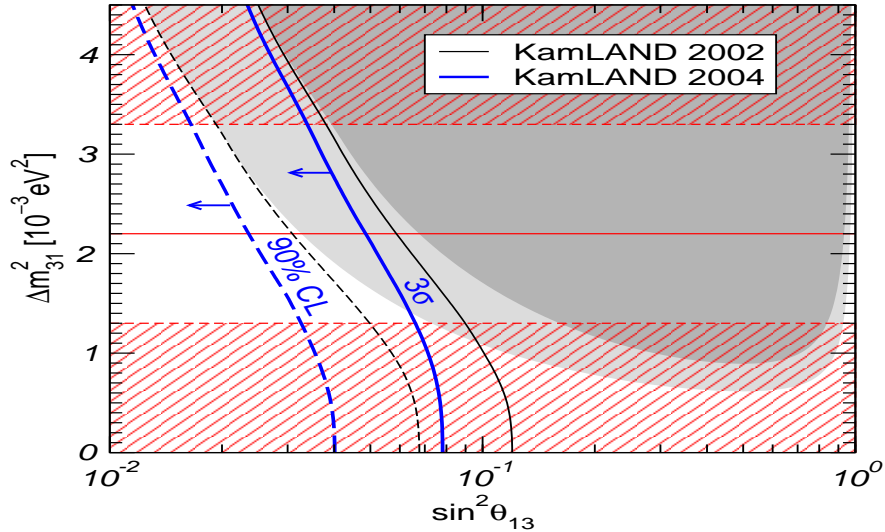}
    \caption{\label{fig:t13-solar-chooz} Upper bound on
      $\sin^2\theta_{13}$ (1 \dof) from solar and reactor data versus
      $\Dma$. Dashed (solid) curves correspond to 90\% (3$\sigma$)
      \CL\ bounds, thick curves include KamLAND-2004 data, thin ones
      do not.  Light (dark) regions are excluded by CHOOZ at 90\%
      (3$\sigma$) \CL\ Hatched regions are excluded by ATM + K2K at
      3$\sigma$, horizontal line corresponds to current $\Dma$ best fit value.}
\end{figure}

In Fig.~\ref{fig:t13-solar-chooz} we show the upper bound on
$\sin^2\theta_{13}$ as a function of $\Dma$ from CHOOZ data alone
compared to the bound from an analysis including solar and reactor
neutrino data. One sees that, although for larger $\Dma$ values the
bound on $\sin^2\theta_{13}$ is dominated by CHOOZ, this bound deteriorates 
quickly as $\Dma$ decreases (see Fig.~\ref{fig:t13-solar-chooz}), so that
for $\Dma \lsim 2 \times 10^{-3} \eVq$ the solar and KamLAND data become 
relevant. 

In summary the overall improvement is especially important for lower $\Dma$ 
values. The new overall global bound on $\sin^2\theta_{13}$ is 0.043 at 
3$\sigma$ for 1 \dof\ Such an improved $\sin^2\theta_{13}$ bound follows 
mainly from the strong spectral distortion found in the 2004 KamLAND sample.

Future long baseline reactor and accelerator neutrino oscillation
searches~\cite{Lindner:2005af}, as well as studies of the day/night
effect in large water Cerenkov solar neutrino experiments such as UNO
or Hyper-K~\cite{SKatm04} could bring more information on
$\sin^2\theta_{13}$~\cite{Akhmedov:2004rq}. With neutrino physics
entering the precision age it is necessary to scrutize also the
validity of the unitary approximation of the lepton mixing matrix in
future experiments, given its theoretical
fragility~\cite{schechter:1980gr}. Indeed, any model where
neutrino masses follow "a-la-seesaw" gives corrections to this
approximation, which may be sizeable in some cases.
                                                                           
\section{Absolute neutrino mass scale}

\begin{figure}[b]
  \centering
\includegraphics[width=6cm,height=4cm]{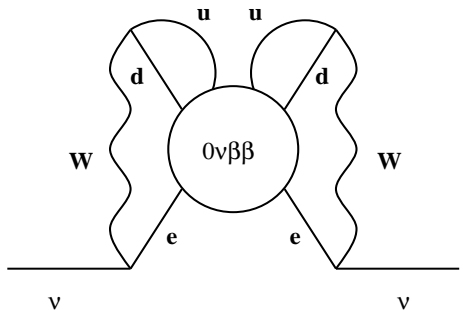}
  \caption{Neutrinoless double beta decay and Majorana mass are theoretically 
  equivalent~\cite{Schechter:1981bd}.}
 \label{fig:bbox}
\end{figure}

Neutrino oscillation data are insensitive to the absolute scale of
neutrino masses and also to the fundamental issue of whether neutrinos
are Dirac or Majorana particles~\cite{schechter:1981gk,doi:1981yb}.
On general grounds neutrino masses are expected to be
Majorana~\cite{schechter:1980gr}, a fact that may explain their
relative smallness with respect to other fermion masses. 
The significance of the neutrinoless double beta decay stems from the fact that, 
in a gauge theory, irrespective of the mechanism that induces \nbb, it necessarily 
implies a Majorana neutrino mass~\cite{Schechter:1981bd}, as illustrated in Fig.
\ref{fig:bbox}. Hence the importance of searching for neutrinoless double beta 
decay~\cite{Wolfenstein:1981rk}. 
Quantitative implications of the ``black-box'' argument are
model-dependent, but the theorem itself holds in any ``natural'' gauge
theory.

Now that oscillations are experimentally confirmed we know that \nbb
must be induced by the exchange of light Majorana neutrinos, the so-called
"mass-mechanism". The corresponding amplitude is sensitive both to the absolute 
scale of neutrino mass as well as the two Majorana CP phases that characterize
the minimal 3-neutrino mixing matrix~\cite{schechter:1980gr}, none of which
can be probed in oscillations.
\begin{figure}[t]
 \centering
\includegraphics[width=.7\linewidth,height=12cm]{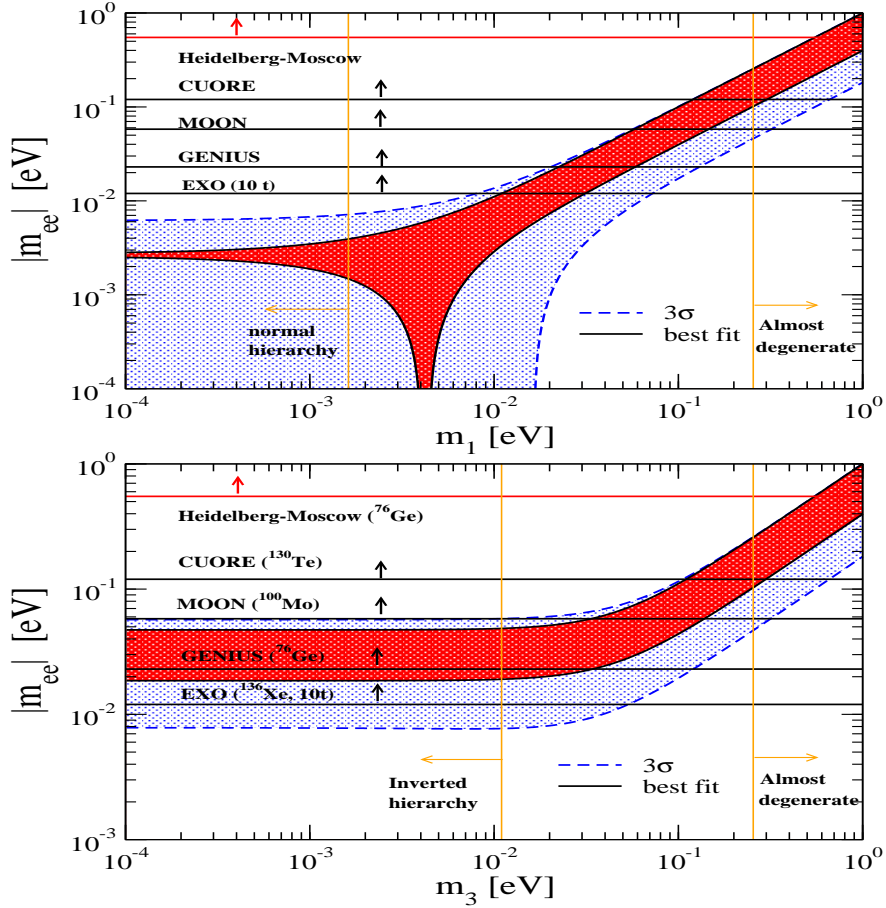}
 \caption{Neutrinoless double 
beta decay amplitude versus current oscillation data.}
\label{fig:nbbfut}
\end{figure}

Fig. \ref{fig:nbbfut} shows the estimated average mass parameter
characterizing the neutrino exchange contribution to \nbb versus the
lightest neutrino mass.  The calculation takes into account the
current neutrino oscillation parameters in \cite{Maltoni:2004ei} and
the nuclear matrix elements of~\cite{Bilenky:2004wn} and compares with
experimental sensitivities.
The upper (lower) panel corresponds to the cases of normal (inverted)
neutrino mass spectra. In these plots the ``diagonals'' correspond to
the case of quasi-degenerate
neutrinos~\cite{caldwell:1993kn,babu:2002dz}, which give the largest
\nbb amplitude.
In the normal hierarchy case there is in general no lower bound on 
the \nbb rate since there can be a destructive interference amongst
the neutrino amplitudes. In contrast,  the inverted neutrino mass
hierarchy implies a ``lower'' bound for the \nbb amplitude.
A normal hierarchy model with no lower bound on \nbb is given in
Ref.~\cite{Hirsch:2005mc}.

Future experiments like GERDA, SuperNEMO, CUORE, COBRA and others will extend 
the sensitivity and provide an independent check of the Heidelberg-Moscow
claim~\cite{Aalseth:2002rf,Klapdor-Kleingrothaus:2004wj}.
More information on the absolute scale of neutrino mass will also come
from future beta decays searches (KATRIN)~\cite{Osipowicz:2001sq} as well 
as cosmology~\cite{Hannestad:2004nb}.
            
\section{The origin of neutrino mass}

It is well-known that the effective dimension-five operator 
$\ell \ell \phi \phi$ where $\phi$ the \21 Higgs doublet and $\ell$ 
is a lepton doublet induces neutrino masses once the electroweak symmetry 
breaks down through a nonzero vacuum expectation value $\vev{\phi}$~\cite{Weinberg:1980bf}.  Nothing is known from first principles about the mechanism giving rise to this operator, 
its associated mass scale or flavour structure.  The resulting Majorana neutrino masses 
are therefore unpredicted in general. However the very fact that Majorana neutrino masses
violate lepton number may explain, irrespective of the underlying physics,
why neutrinos are much lighter than the other fermions.

One possibility is that the dimension-five operator is suppressed by a large 
scale in the denominator (top-bottom scenario). Alternatively, it may be 
suppressed by a small scale in the numerator (bottom-up scenario).
Both scenarios are viable and can be made natural, the first being closer
to the idea of unification.

The most well-studied realization of the top-bottom scenario is the seesaw 
mechanism, which induces small neutrino masses from the exchange of heavy 
states that might come from unification. 
\begin{figure}[t] \centering
    \includegraphics[height=3cm,width=.45\linewidth]{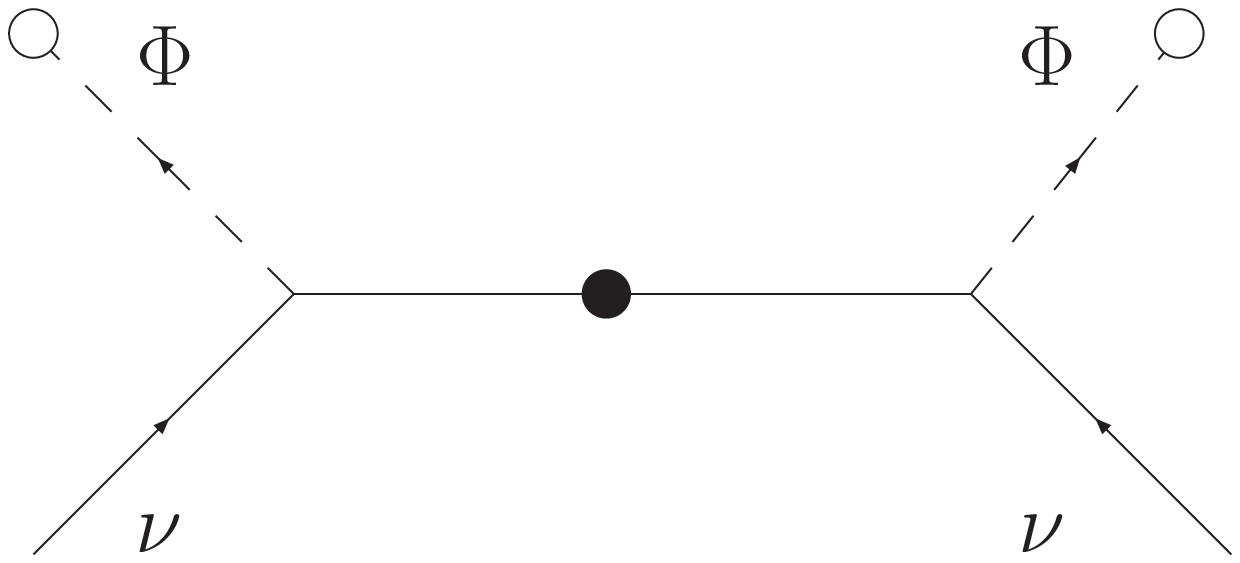}
    \includegraphics[height=3.2cm,width=.45\linewidth]{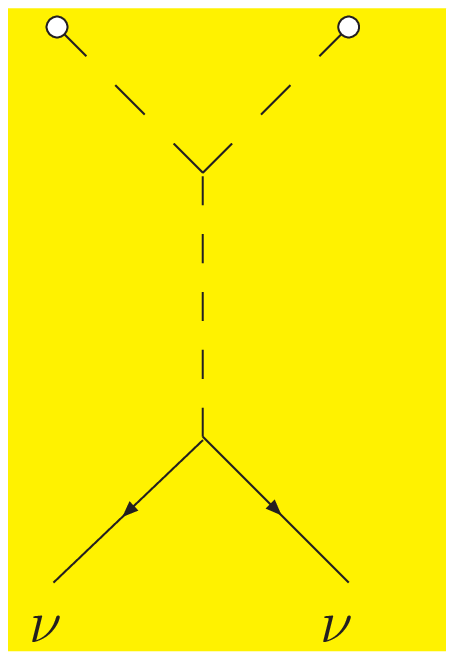}
    \caption{\label{fig:seesaw} %
     Two types of seesaw mechanism.}
\end{figure}
Small neutrino masses are induced either by heavy \21 singlet
``right-handed'' neutrino exchange (type I) or heavy scalar bosons
exchange (type II), in a nomenclature opposite from the one given
in~\cite{schechter:1980gr}. The effective triplet seesaw term has a
flavor structure different from the type-I term, contributing to the
lack of predictivity of general seesaw schemes, where both co-exist. 
Predicitvity within the seesaw approach may be obtained by appealling 
to extra symmetries, as given, for example, in~\cite{babu:2002dz}. 
The model predicts maximal $\theta_{23}=\pi/4$, $\theta_{13}=0$, and 
$\theta_{12}=\cal O$(1), though unpredicted. Moreover, if CP is violated 
$\theta_{13}$ becomes arbitrary but the Dirac CP violation phase is 
maximal~\cite{Grimus:2003yn}. The model leads to a variety of phenomenological 
implications. For example, it gives a lower bound on the absolute neutrino 
mass $m_{\nu}\gsim 0.3$ eV. It also requires a light slepton below 200 GeV, 
and gives large rates for flavour violating processes. A survey of related 
models is given in Ref.~\cite{mnuth}.

Amongst ``bottom-up'' models we mention those where neutrino masses
are given as radiative corrections~\cite{zee:1980ai} and
models where low energy supersymmetry is the origin of neutrino
mass~\cite{Hirsch:2004he}.  The latter are based on the idea that R
parity spontaneously breaks~\cite{Masiero:1990uj}, leading to a very
simple effective bilinear R parity violation model~\cite{Diaz:1997xc}.
In this case the neutrino mass spectrum typically follows a normal hierarchy, 
with the atmospheric scale generated at the tree level and the solar scale
radiatively ``calculable''~\cite{Hirsch:2000ef}. In order to reproduce the 
masses indicated by current data, typically the lightest supersymmetric 
particle decays inside the detector. More strikingly, its decay properties 
correlate with neutrino mixing angles. For example, if the LSP is the 
lightest neutralino, it is expexted to decay 50/50 to muons and taus, 
since the observed atmospheric angle is close to $\pi/4$~\cite{Hirsch:2004he,Hirsch:2000ef}.
This constitutes a characteristic feature of the proposal that
supersymmetry is the origin of neutrino mass and opens the tantalizing
possibility of testing neutrino mixing at high energy accelerators,
like the "Large Hadron Collider" (LHC) and the "International Linear
Collider" (ILC).

\section{Robustness of the oscillation interpretation}

The general effective model-independent description of the seesaw at
low-energies is characterized by $(n,m)$, $n$ being the number of \21 
isodoublet and $m$ the number of \21 isosinglet leptons~\cite{schechter:1980gr}.
This leads to a very rich and complex structure of the charged current 
lepton mixing matrix (non-unitary) and non--diagonal neutral current~\cite{schechter:1980gr}. For example, the $(3,3)$ seesaw model has 12 mixing angles and 12 CP phases 
(both Dirac and Majorana-type) characterizing its full 3$\times$6 lepton mixing 
matrix~\cite{schechter:1980gr}.

\begin{figure}[t] \centering
    \includegraphics[height=3.7cm,width=.5\linewidth]{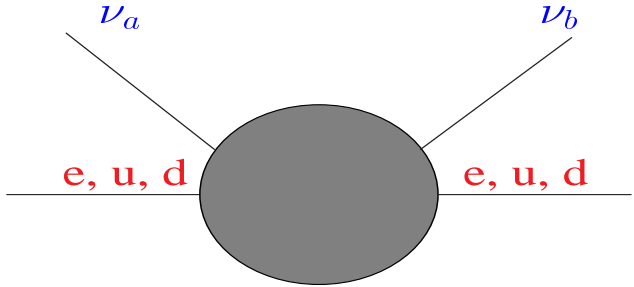}
    \caption{\label{fig:nuNSI} %
      Flavour-changing effective operator for non-standard neutrino interaction.}
\end{figure}

The nontrivial structure of charged and neutral current weak
interactions is a general feature of seesaw models~\cite{schechter:1980gr} 
and leads to dimension-6 terms non-standard neutrino interactions (NSI), as
illustrated in Fig.~\ref{fig:nuNSI}. 
Such sub-weak strength $\varepsilon G_F$ operators can be of two
types: flavour-changing (FC) and non-universal (NU). 
In inverse seesaw-type models~\cite{mohapatra:1986bd} the non-unitary
piece of the lepton mixing matrix can be sizeable and hence the induced
non-standard interactions may be phenomenologically important~\cite{bernabeu:1987gr}.
"Large" NSI strengths may also be induced by scalar boson 
exchanges in models with radiatively induced neutrino masses~\cite{zee:1980ai},
and in supersymmetric unified models~\cite{hall:1986dx}.
                                                                
Non-standard physics may in principle affect neutrino propagation
properties and detection cross sections~\cite{pakvasa:2003zv}.
In their presence, the Hamiltonian describing atmospheric neutrino
propagation has, in addition to the standard oscillation part, another
term $H_\mathrm{NSI}$ 
\begin{equation}
    H_\mathrm{NSI} = \pm \sqrt{2} G_F N_f
    \left( \begin{array}{cc}
        0 & \varepsilon \\ \varepsilon & \varepsilon'
    \end{array}\right) \,.
\end{equation}
Here $+(-)$ holds for neutrinos (anti-neutrinos) and $\varepsilon$ and
$\varepsilon'$ parameterize the NSI: $\sqrt{2} G_F N_f \varepsilon$ is
the forward scattering amplitude for the FC process $\nu_\mu + f \to
\nu_\tau + f$ and $\sqrt{2} G_F N_f \varepsilon'$ represents the
difference between $\nu_\mu + f$ and $\nu_\tau + f$ elastic forward
scattering. Here $N_f$ is the number density of the fermion $f$ along
the neutrino path.  In the 2--neutrino approximation, the
determination of atmospheric neutrino parameters $\Dma$ and
$\sin^2\theta_\Atm$ was shown to be practically unaffected by the
presence of NSI on down-type quarks ($f=d$)~\cite{fornengo:2001pm}.
Future neutrino factories will substantially improve this
bound~\cite{huber:2001zw}. 

In contrast, the oscillation interpretation of solar neutrino data is
more ``fragile'' with respect to the presence of non-standard
interactions in the $e-\tau$ sector~\cite{Miranda:2004nb}.
On the other hand, it has been shown~\cite{huber:2002bi} that, even a
small residual non-standard interaction of neutrinos in the $e-\tau$
channel leads to a drastic loss in sensitivity in the $\theta_{13}$
determination at a neutrino factory. It is therefore important to
improve the sensitivities on NSI, another window of oportunity for
neutrino physics in the precision age.
   
\vskip .5cm 
Work supported by Spanish grants FPA2005-01269 and BFM2002-00345 and by 
the EC RTN network MRTN-CT-2004-503369. 

\def\baselinestretch{1.1}


\end{document}